\documentclass[prl,twocolumn,showpacs,a4paper]{revtex4-1}
\usepackage{epsfig}
\usepackage{amsmath}
\usepackage{graphicx}
\usepackage{color}
\usepackage{hyperref}
\usepackage{bbm}

\newcommand{\di}{\partial}

\newcommand{\ket}[1]{\vert{#1}\rangle}

\newcommand{\abs}[1]{\left|#1\right|} 

\newcommand{\beq}{\begin{equation}}
\newcommand{\eeq}{\end{equation}}

\newcommand{\be}{\begin{equation}}
\newcommand{\ee}{\end{equation}}

\newcommand{\ben}{\begin{eqnarray}}
\newcommand{\een}{\end{eqnarray}}
\usepackage{amsmath}

\begin{document}

\title{
Probing macroscopic realism via Ramsey correlations measurements
}

\author{ A. Asadian$^{1}$, C. Brukner$^{2,3}$, P. Rabl$^1$}
\affiliation{$^1$Institute of Atomic and Subatomic Physics, TU Wien, Stadionallee 2, 1020 Wien, Austria}
\affiliation{$^2$Faculty of Physics, University of Vienna, Boltzmanngasse 5, 1090 Vienna, Austria}
\affiliation{$^3$Institute for Quantum Optics and Quantum Information,
Austrian Academy of Sciences, Boltzmanngasse 3, 1090 Vienna, Austria}

\date{\today}

\begin{abstract}
%
We describe a new and experimentally feasible protocol for performing fundamental tests of quantum mechanics with massive objects. In our approach a single two level system is used to probe the motion of a nanomechanical resonator via multiple Ramsey interference measurements. This scheme enables the measurement of \emph{modular variables} of macroscopic continuous variable systems and we show that  correlations thereof violate a Leggett-Garg inequality and can be applied for tests of quantum contextuality. Our method can be implemented with a variety of different solid state or photonic qubit-resonator systems and provides  a clear experimental signature to distinguish the predictions of quantum mechanics from those of other alternative theories at a macroscopic scale. 
\end{abstract}


\maketitle

In his celebrated paper in 1964 Bell showed that the laws of quantum mechanics are inconsistent with a description of our world based on local elements of reality \cite{Bell}.  
Bell derived an experimentally testable inequality, which bounds the correlations between bipartite measurements for \emph{any} local hidden variable theory, but which is violated by quantum mechanics. Since then the results constraining the permissible types of hidden variable models of quantum mechanics have attracted much attention and have been reformulated as the problem of contextual measurements by Kochen and Specker~\cite{Kochen} and in terms of temporal correlations by Leggett and Garg~\cite{LG85}. Today these concepts have been tested in various experiments with photons~\cite{Aspect1999}, ions~\cite{Roos09}, impurity spins~\cite{WaldherrPRL2011,KneeNatComm2012} or superconducting qubits~\cite{Palacios-LaloyNatPhys2010,GroenPRL2013} confirming quantum mechanics on a microscopic level. The challenge is now to verify or disprove these predictions also with more massive objects~\cite{OxfordQuestions}, where quantum physics conflicts with our daily life perceptions as well as with alternative theories and (gravity-induced) collapse models \cite{Diosi,Penrose,GPW,GhirardiPRA1990,NimmrichterPRA2011,PikovskiNatPhys2012,WenzelProcRSocA2012,YangPRL2013}.

\begin{figure}[b]
\begin{center}
\includegraphics[width=0.48\textwidth]{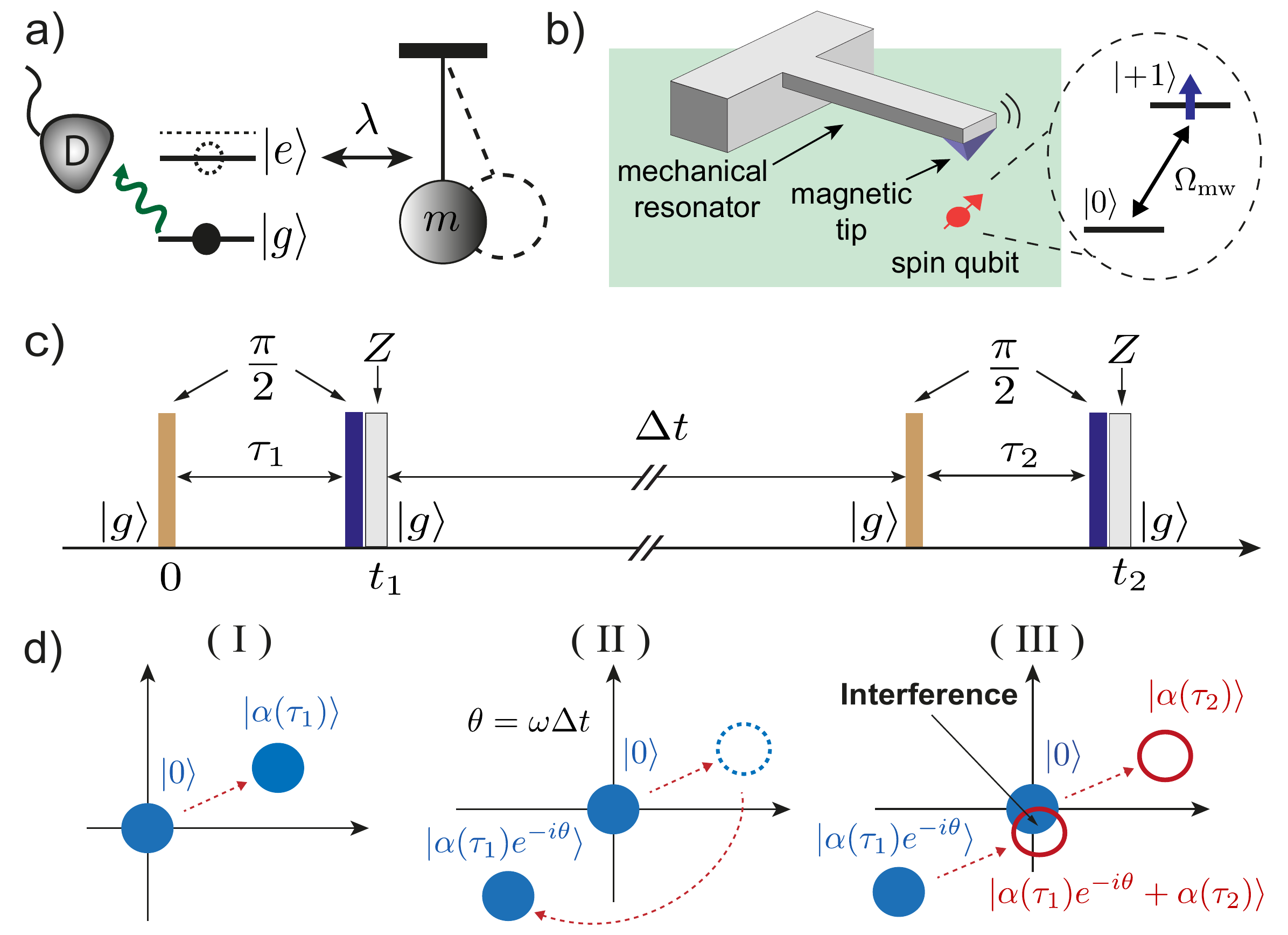}
\caption{(color online). a) Schematic setup. The motion of a macroscopic mechanical resonator modulates the excited state energy of a two level system with strength $\lambda$. b) This system can be implemented, for example, by coupling a magnetic tip to the two spin states $|g\rangle\equiv|m_s=0\rangle$ and $|e\rangle\equiv|m_s=+1\rangle$ of a nitrogen-vacancy (NV) center in diamond~\cite{RablPRB2009}.
c) Pulse sequence for two RMs separated by a waiting time $\Delta t$. Each RM consists of two $\pi/2$ pulses separated by a variable interaction time $\tau_i$, and followed by a projective measurement on the $Z$ basis,  $\{ \ket{g},\ket{e}\}$. Before and after each Ramsey sequence the qubit is initialized in state $|g\rangle$. d) Illustration of the conditioned resonator superposition state (I) after the first measurement, (II) after the waiting period and (III) at the end of the second Ramsey sequence. 
}
\label{fig:Setup}
\end{center} 
\end{figure}

In recent years a rapid progress towards the quantum control of nano- and micromechanical systems has been achieved: resonators with masses in the picogram regime have been cooled close to the quantum ground state~\cite{OConnellNature2010,TeufelNature2011,ChanNature2011,VerhagenNature2012} and first steps for coupling mechanical resonators to single electronic spins~\cite{ArcizetNatPhys2011,KolkowitzScience2012} or superconducting qubits \cite{LaHayeNature2009,OConnellNature2010,PirkkalainenNature2013} have been implemented. In this work we show, how these techniques can be directly applied for testing the most fundamental aspects of quantum mechanics on a macroscopic scale. The general idea is illustrated in Fig.~\ref{fig:Setup}, where a microscopic two level system (qubit) is coupled to a massive mechanical resonator and is used to probe the resonator displacement via multiple Ramsey measurements (RMs) ~\cite{KolkowitzScience2012,BennettNJP2012,FinkPRL2013}.  Our analysis shows that the correlations between two subsequent RMs can violate a Leggett-Garg inequality (LGI)~\cite{LG85,Nori}, and thereby provide a clear experimental signature for distinguishing the predictions of quantum mechanics from those of other realistic theories. 
More generally, our scheme allows the measurement of so-called \emph{modular variables}, which, for example, play an important role for the  detection of non-local phases~\cite{Aharonov,Popescu2010} or tests of quantum contextuality in continuous variable systems~\cite{PlastinoPRA2010}.  
This makes it a versatile tool for various tests of quantum mechanics, which extend previous ideas~\cite{LambertPRB2011} to the full Hilbert space of a collective macroscopic variable and complement quantum interference studies with large photonic states~\cite{LvovskyNatPhys2013,BrunoNatPhys2013} and truly massive objects~\cite{BosePRA1999,ArmourPRL2002,TianPRA2005,MarshallPRL2003,RomeroIsartPRL2011,NimmrichterPRA2011,WenzelProcRSocA2012}.


\emph{Model}. We consider the general setup depicted in Fig.~\ref{fig:Setup} a), where a mechanical resonator with oscillation frequency $\omega$ is coupled to a qubit with states $|g\rangle$ and $|e\rangle$, via magnetic~\cite{RablPRB2009}, electrostatic~\cite{ArmourPRL2002,TianPRA2005} or radiation pressure~\cite{MarshallPRL2003,VacantiPRA2013,Akram2013} interactions. The qubit states are split by a large frequency $\omega_q$, which is modulated by the vibrations of the resonator. 
The system is described by the Hamiltonian $(\hbar=1)$,
\begin{equation} \label{eq:Hamiltonian} 
H= \omega_q |e\rangle\langle e| +  \omega  a^\dag a   + \lambda (a+a^\dag)|e\rangle\langle e|, 
\end{equation}
where $a$ ($a^\dag$) is the annihilation (creation) operator of the mechanical mode and $\lambda$ is the coupling strength. Note that the bare qubit Hamiltonian $H_q=\omega_q|e\rangle\langle e|$ commutes with the interaction and in the following we will work in a rotating frame, where this term is omitted.
%
%
Eq.~\eqref{eq:Hamiltonian} describes a frequency shift of the state $|e\rangle$ proportional to the resonator displacement, $x=(a+a^\dag)/\sqrt{2}$, which can be detected by performing a sensitive RM of the qubit transition frequency~\cite{KolkowitzScience2012,BennettNJP2012}. 
Here we are interested in the non-classical correlations, which arise from the associated quantum backaction in a sequence of two or multiple RMs [cf. Fig.~\ref{fig:Setup} c)].

\emph{Ramsey measurements and  modular variables.}
For the implementation of a single RM the qubit is initially prepared in state $|g\rangle$ and rotated at time $t=0$ 
into a superposition state $R_{\pi/2}(\varphi_1)|g\rangle=(|g\rangle +e^{i\varphi_1}|e\rangle)/\sqrt{2}$ by applying a fast $\pi/2$ pulse. The system then evolves under the action of Hamiltonian~\eqref{eq:Hamiltonian} for a time $\tau_1$, which creates a state-dependent displacement of the resonator mode and entangles the spin and the mechanical degrees of freedom. For example, for a resonator, which is initially prepared in the ground state $|0\rangle$, the systems evolves into the superposition state $ (|0\rangle|g\rangle + e^{i \bar \varphi_1} |\alpha(\tau_1)\rangle|e\rangle)/\sqrt{2}$, where 
$|\alpha(\tau_1)\rangle$ is a coherent state with amplitude $\alpha(\tau)=\lambda/\omega(e^{-i\omega \tau}-1)$ and $\bar \varphi_i= \varphi_i+\phi(\tau_i)$ includes an additional geometric phase $\phi(\tau)=\lambda^2/\omega^2(\omega\tau-\sin \omega\tau )$. Finally, a second $\pi/2$ pulse, $R_{\pi/2}(0)$, is applied and the state of the qubit is measured. 
For an arbitrary initial resonator state $|\psi\rangle$ the evolution generated by the whole pulse sequence, $U_M(\varphi_1,\tau_1)= R_{\pi/2}(0) e^{-iH \tau_1} R_{\pi/2}(\varphi_1)$, is 
\begin{equation}\label{eq:UM}
U_M(\varphi_1,\tau_1)|\psi\rangle|g\rangle = E_-(\varphi_1,\tau_1) |\psi\rangle |g\rangle+ E_+ (\varphi_1,\tau_1) |\psi\rangle |e\rangle.
\end{equation}
Here  $E_\pm(\varphi,\tau) =\frac{1}{2}\left[\mathbbm{1} \pm e^{i \bar \varphi(\tau)}\mathcal{D}(\alpha(\tau)) \right]ÊU_0(\tau)$ are Kraus operators satisfying $E_+^\dag E_++E_-^\dag E_-=\mathbbm{1}$,  $\mathcal{D}(\alpha)=e^{\alpha a^\dag-\alpha a}$ is the displacement operator and $U_0(t)= e^{-i\omega t a^\dag a}$. Eq.~\eqref{eq:UM} can be readily generalized to arbitrary (mixed) initial resonator states $\rho_0$~\cite{SuppInfo} and the resulting probabilities for finding the qubit at time $t_1=\tau_1$ in state $|g\rangle$ ($p_-$) or state $|e\rangle$ ($p_+$) are $p_\pm={\rm Tr}\{ÊE_\pm^\dag(\varphi_1,\tau_1) E_\pm(\varphi_1,\tau_1) \rho_0\}$. For a resonator, which is initially prepared in the ground state, $p_\pm=(1\pm \cos(\bar \varphi_1)e^{-|\alpha(\tau_1)|^2/2})/2$. In previous works, 
the resulting collapse and revivals of $p_\pm$ as a function of $\tau_1$ have been suggested as a way to probe the mechanical superpositions, which are generated during the interaction time \cite{ArmourPRL2002,TianPRA2005,MarshallPRL2003}. However, similar collapse and revival signals also occur for a purely classical resonator~\cite{KolkowitzScience2012,BennettNJP2012, SuppInfo} and are not a conclusive signature for quantum behavior on their own.

In the following discussion we will extend the above analysis to multiple RMs and consider correlations between the dichotomic variables $Z(t_n)=\pm 1$, which describe the outcome of the $n$-th measurement at time $t_n$. For the first measurement we obtain
\begin{equation}\label{eq:Z1}
\langle Z(t_1)\rangle= p_+-p_- = {\rm Tr}\{ÊQ(\bar \varphi_1,\alpha_{1}) \rho_0 \},
\end{equation} 
where $Q(\varphi,\alpha)=\frac{1}{2}[e^{-i \varphi} \mathcal{D}(-\alpha)+e^{i \varphi} \mathcal{D}(\alpha)]$ and we have introduced the abbreviation $\alpha_{n}\equiv \alpha (\tau_n)e^{i\omega t_n}$. 
The operator appearing in Eq.~\eqref{eq:Z1} can further be written as
\begin{equation}
Q(\varphi,\alpha)=\text{cos}\left(\varphi+\sqrt{2}{\rm Im}(\alpha)x -\sqrt{2} {\rm Re}(\alpha)p\right),
\end{equation}
where 
$p=i(a^\dag-a)/\sqrt{2}$. Observables of this type are known as \emph{modular variables} and 
play an important role for extending fundamental concepts in quantum mechanics from two-level to continuous variable systems. General $n$-point correlation functions $C(t_1,t_2,\dots t_n)= \langle Z(t_n)\dots Z(t_2)Z(t_1)\rangle $ are given by~\cite{SuppInfo}  
\begin{equation}\label{eq:Correlations}
C(t_1,t_2,\dots t_n) = {\rm Tr}\{\mathcal{Q}_{t_n} \dots  \mathcal{Q}_{t_2}\mathcal{Q}_{t_1} \rho_0 \},
\end{equation} 
where the superoperator $\mathcal{Q}_{t_n}$ is defined by $\mathcal{Q}_{t_n} \rho= [e^{i \bar \varphi_n} \mathcal{D}(\alpha_{n}) \rho + \rho e^{-i \bar \varphi_n} \mathcal{D}^\dag(\alpha_{n})]/2$. For multiple RMs the different displacements induced by each measurement in general no longer commute. Therefore, the resulting correlation functions differ from the corresponding classical results, as we show in more detail below.

\emph{Probing quantum superpositions.}  Eq.~\eqref{eq:UM} shows that conditioned on the outcome of $Z(t_1)$ the resonator is projected into one of the states $ |\psi^\pm\rangle=E_\pm(\varphi_1,\tau_1)|\psi\rangle/\sqrt{p_\pm}$, which for a resonator initially in the ground state $|\psi\rangle=|0\rangle$, are explicitly given by~\cite{TianPRA2005,SteinkePRA2011,BennettNJP2012,VacantiPRA2013}
\begin{equation}\label{eq:Superposition1}
 |\psi^\pm\rangle = \frac{|0\rangle \pm e^{i \bar \varphi_1} |\alpha(\tau_1)\rangle}{\sqrt{4p_\pm}}. 
\end{equation}
Modular variables have first been introduced for detecting the non-local phase $\bar \varphi_1$ of such a spatially separated superposition state~\cite{Aharonov,Popescu2010}. Following this idea, we now show that the expectation value $\langle Z(t_2)\rangle$ of a second RM can be used to probe the coherence of the macroscopic superposition generated by the first measurement.

After the measurement the qubit is reset to state $|g\rangle$ and the state in Eq.~\eqref{eq:Superposition1} evolves freely for a time $\Delta t$ during which the qubit is decoupled from the resonator.
%
%
At time $t=t_2-\tau_2$ a second Ramsey sequence with duration $\tau_2$, amplitude $\alpha(\tau_2)$ and phase $\bar \varphi_2=\varphi_2+\phi(\tau_2)$ is applied. By 
assuming the outcome $Z(t_1)=+1$, the state immediately before the second qubit readout is  $U_M(\varphi_2,\tau_2)U_0(\Delta t)|\psi^+\rangle|g\rangle= |\psi^{-|+}\rangle|g\rangle+ |\psi^{+|+}\rangle|e\rangle$, where (up to a global rotation in phase space)
\begin{align}\label{eq:4states}
|\psi^{\pm|+}\rangle=\frac{|0\rangle\!+\!e^{i\bar \varphi_1}|\alpha_{1}\rangle\!\pm \!e^{i\bar \varphi_2}|\alpha_{2}\rangle\!\pm\! e^{i (\bar \varphi_{1}+ \bar \varphi_{1}+\gamma)} |\alpha_{1}\!+\!\alpha_{2}\rangle}{4\sqrt{p_+}}.
\end{align}
In this expression $\gamma={\rm Im}\{Ê\alpha_1^*\alpha_2\}$ is an additional phase, which arises from the two non-commuting displacements in the first and the second measurement. 
From Eq.~\eqref{eq:4states} we obtain the conditioned probabilities $p_{\eta_2|\eta_1}$ for finding, for example, the qubit in state $|g\rangle$ ($\eta_2=-$) given that in the first measurement the qubit was found in state $|e\rangle$ ($\eta_1=+$). As illustrated in Fig.~\ref{fig:Setup} d), these probabilities now depend on the interference between the first superposition and the displacement $\alpha_2$ generated in the second pulse sequence.
%
%

\begin{figure}
  \begin{center}
 \includegraphics[width=0.48\textwidth]{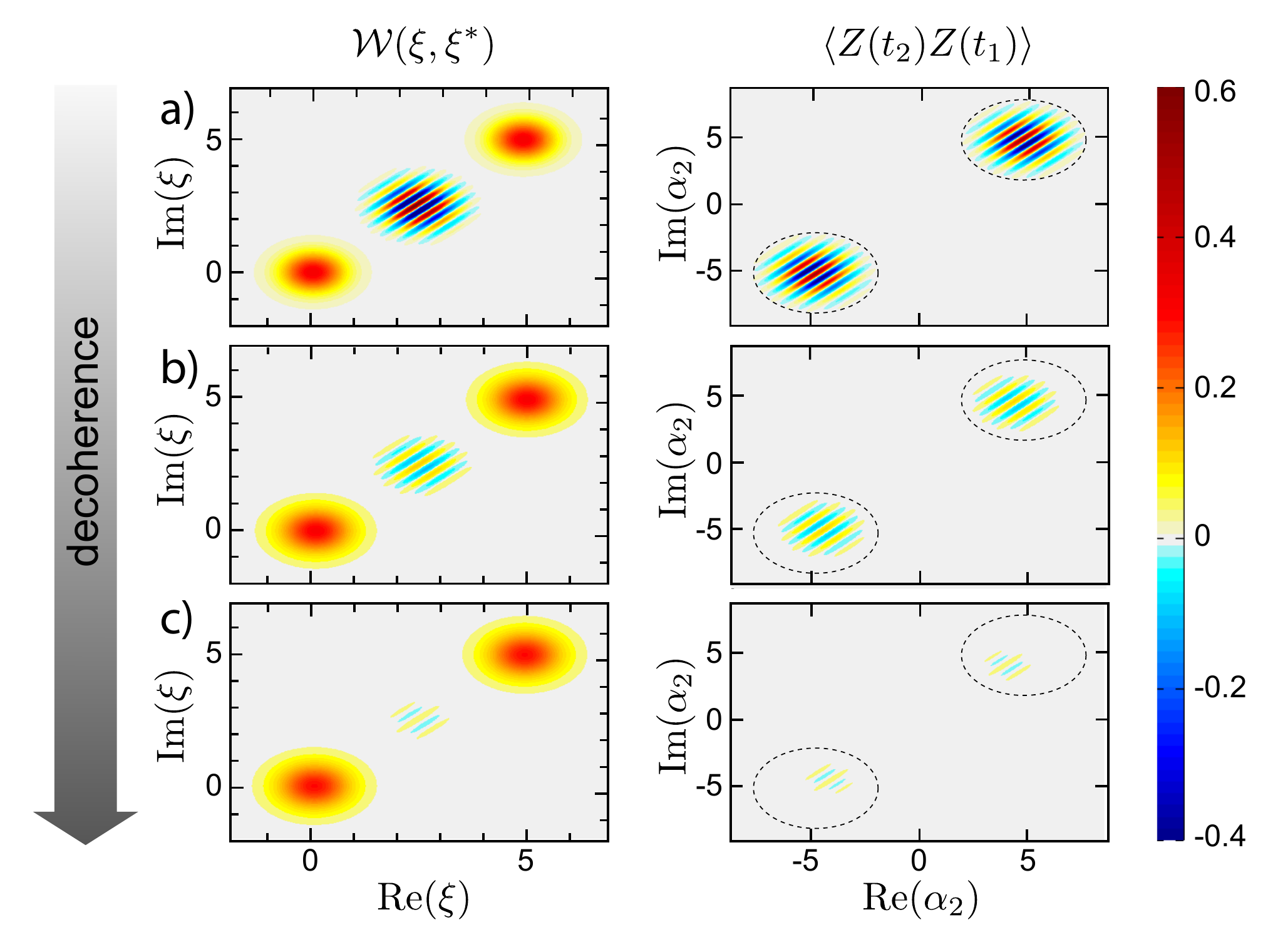} 
 \caption{The Wigner function $\mathcal{W}(\xi,\xi^*)$ (left column) corresponding to the conditioned resonator state $|\psi^+\rangle= (|0\rangle+ e^{i \bar \varphi_1} |\alpha_1)\rangle)/\sqrt{4p_+}$, is compared with the correlation $\langle Z(t_2)Z(t_1)\rangle$ (right column) between two RMs. In a) a fully coherent evolultion and $\bar \varphi_{1,2}=0$ and $\alpha_1=5+5i$ has been assumed. In the lower two rows the effect of mechanical decoherence with rates b) $\Gamma_{\rm th}\Delta t=0.04$ and c) $\Gamma_{\rm th}\Delta t=0.08$ during the free evolution is taken into account  (see text and \cite{SuppInfo} for more details).}  
 \label{fig:Interference}
 \end{center}
 \end{figure}

In Fig.~\ref{fig:Interference} we plot the Wigner function of the conditioned state $|\psi^+\rangle$ together with the resulting correlation $\langle Z(t_2)Z(t_1)\rangle=p_+(p_{+|+}- p_{-|+})+p_-(p_{-|-}-p_{+|-})$ 
These correlations vanish almost everywhere, except for $ \alpha_2 \approx \pm \alpha_1$, where two parts of the superposition state given in Eq.~\eqref{eq:4states} interfere.  
Fig.~\ref{fig:Interference} b) and c) show the outcome of the same measurement, but assuming that the mechanical state is subject to decoherence during the free evolution time $\Delta t$. 
We model mechanical dissipation by a master equation with a characteristic thermal decoherence rate $\Gamma_{\rm th}\simeq k_B T/\hbar Q$~\cite{SuppInfo}, where $T$ is the support temperature and $Q$ the mechanical quality factor. As the resonator state evolves from a coherent superposition into a classical mixture $\simeq \frac{1}{2}(|0\rangle\langle 0|+|\alpha_1\rangle\langle \alpha_1|)$ (indicated by the decay of the fringes of the Wigner function), the contrast of the correlation $\langle Z(t_2)Z(t_1)\rangle$ degrades and eventually vanishes completely.  This illustrates, how the Ramsey correlation method can be used to simultaneously prepare and probe the survival of quantum superposition states.  Since during the waiting time $\Delta t$ the mechanical superposition is completely decoupled from decoherence mechanisms affecting the qubit, this method is particularly suited for high-Q resonators and levitated objects~\cite{Chang2010,RomeroIsartPRL2011,RomeroIsartPRL2012,CirioPRL2012,Yin2013,Yin2013b}.

\emph{Leggett-Garg inequality.} Following Leggett \cite{Leggett}, the observation of distinct quantum superpositions is only a first, but still insufficient step to exclude a realistic picture at the macroscopic level. Macrorealism is defined by the conjunction of two essential postulates~\cite{Leggett}: ``(i) Macrorealism per se. A macroscopic object which has available to it two or more macroscopically distinct states is at any given time in a definite one of those states. (ii) Non-invasive measurability.  It is possible in principle to determine
which of these states the system is in without any effect on the state itself or on the subsequent system dynamics." These assumptions allow the derivation of LGIs  
of the type~\cite{Nori},
 \begin{equation}\label{eq:LGinequality}
W= C(t_1,t_2)+ C(t_2,t_3)- C(t_1,t_3) \leq 1,
\end{equation}
which impose a bound on the correlations $C(t_i,t_j)$ between variables with values of modulo smaller or equal to 1, which are
measured at different times $t_i<t_j$. 
The present technique allows us to
measure the correlations between continuous modular observables of the resonator by measuring the correlations between the dichotomic observables $Z(t_i)$ of the qubit.


According to the general result given in Eq.~\eqref{eq:Correlations}, the correlations $C(t_1,t_2)$ between two subsequent RMs are \begin{equation}\label{Qcorr}
\begin{split}
C(t_1,t_2)= \frac{1}{2}\Big[&\cos(\bar \varphi_1+\bar \varphi_2+\gamma) \langle \mathcal{D}(\alpha_1 + \alpha_2)\rangle \\
& +\cos(\bar \varphi_1-\bar \varphi_2-\gamma) \langle \mathcal{D}(\alpha_2 - \alpha_1)\rangle  \Big].
\end{split}
\end{equation}
For a resonator, which is initially prepared in a thermal state with mean occupation number $\bar n$ (which can be lower than the equilibrium occupation number, when the resonator is actively cooled), and assuming identical measurements $
\alpha(\tau_1)=\alpha(\tau_2)=\alpha$, we obtain
\begin{equation}\label{eq:Cij}
\begin{split}
C(t_i,t_j) =
\\  \frac{1}{2} \Big[ \cos(\bar \varphi_i+\bar \varphi_j+\gamma_{ij})& e^{-|\alpha|^2(2\bar n+1)(1+\cos\theta_{ij})} \\
+ \cos(\bar \varphi_i-\bar \varphi_j-\gamma_{ij} )& e^{-|\alpha|^2(2\bar n+1)(1-\cos\theta_{ij})} \Big],
\end{split}
\end{equation}
where $\theta_{ij}=\omega(t_j-t_i)$ and $\gamma_{ij}=|\alpha|^2\sin(\theta_{ij})$. In the following we restrict our discussion to equidistant time intervals, where $\theta_{12}=\theta_{23}=\theta$ and $\theta_{13}=2\theta $.

\begin{figure}
 \begin{center}
\includegraphics[width=0.48\textwidth]{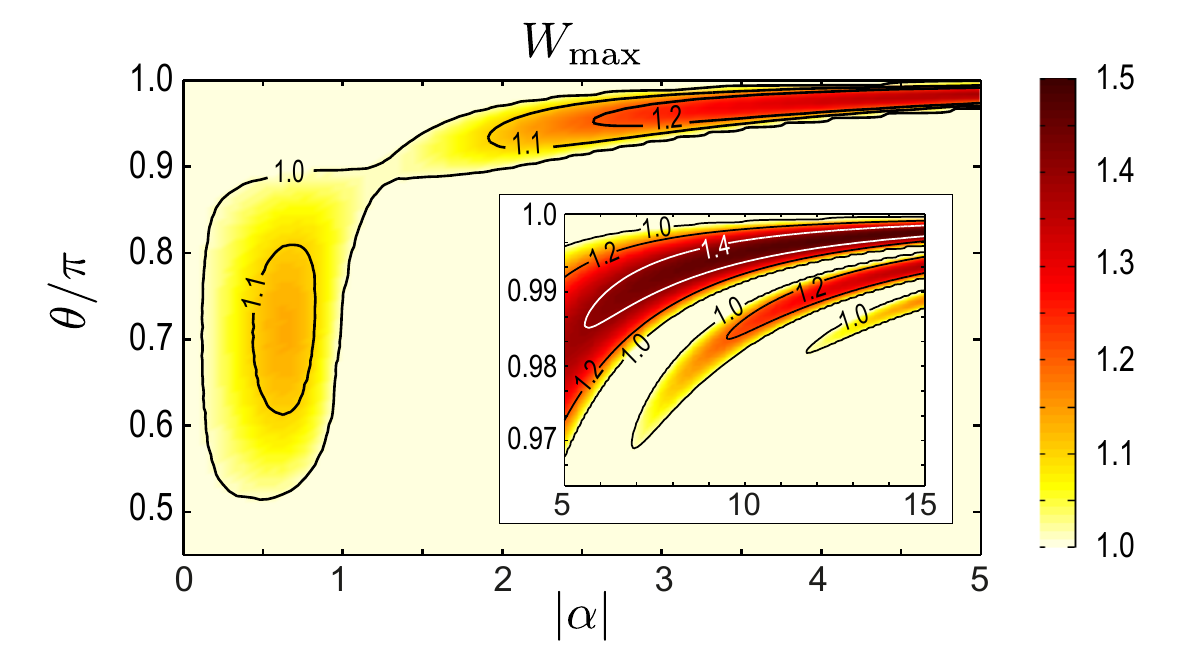} 
\caption{(colour online). The maximal value $W_{\rm max}={\rm max}\{ÊWÊ| \bar \varphi_i \in [0,2\pi]\}$ of the sum of the correlation functions appearing on the left hand side of the LGI~\eqref{eq:LGinequality} is plotted as a function of $\alpha=\alpha(\tau_{1})=\alpha(\tau_2)$ and the free rotation angle $\theta=\omega(t_2-t_1)$. }
\label{fig:LGViolation}
\end{center}
\end{figure}

In Fig.~\ref{fig:LGViolation} we plot the value of $W$ as a function of $\alpha$ and $\theta$ and maximized with respect to the three adjustable phases $\bar \varphi_i$. We see that a significant violation of the LGI already occurs for $\alpha\sim0.5$, where there is still a considerable overlap between the displaced states. For small $\alpha$ 
the violation of the LG inequality is maximal for $\bar \varphi_1=\bar \varphi_2=\pi$ and $\bar \varphi_3=\pi/2$, where $W\simeq 1-|\alpha|^2(2\bar{n}+1)+|\alpha|^2(\sin\theta-\sin 2\theta)$. For $\theta=3\pi/4$ we obtain
$W\simeq 1+|\alpha|^2 (\sqrt{2}/2-2\bar{n})$.
This shows that in principle it is possible to obtain nonclassical correlations for arbitrary small $\alpha$, but the violation is restricted to very small thermal occupancies $\bar n\leq 0.35 $.

For $|\alpha|\gg1$ the correlation function is essentially zero except for $\theta \approx \pi$, where two parts of the four partite superposition given in Eq.~\eqref{eq:4states} interfere [see Fig.~\ref{fig:Setup} d)]. In this limit the optimal choice is $\bar \varphi_1+\bar \varphi_2=-\pi/2$, $\bar \varphi_3=\bar \varphi_1$ and $\theta\approx \pi- \pi/(2|\alpha|^2)$.  In this case
\begin{equation}
W\simeq \frac{3}{2}\left(1-\frac{\pi^2}{4|\alpha|^2}(2\bar n+1)\right),
\end{equation} 
and an almost maximal violation $W=1.5$~\cite{MaxBound} can be achieved for  $|\alpha|\gg1$. The violation is also robust with respect to a finite thermal occupancy, which can be compensated by increasing $|\alpha|$, i.e. by displacing the oscillator state by more than its thermal width. 
%
%
While for a static coupling, $|\alpha|\leq 2\lambda/\omega$, and for many implementations $\lambda/\omega<1$, the displacement amplitude can be increased by periodically flipping the qubit state during the measurement~\cite{TianPRA2005,BennettNJP2012,SuppInfo}. This leads to a parametric amplification of the displacement amplitude up to maximal value $|\alpha_{\rm max}| \leq \frac{\lambda}{\pi} \times {\rm min}\{ÊT_2, T_{\rm th}\}$,  which is limited by the decoherence time $T_2$ of the qubit and the mechanical rethermalization time $T_{\rm th}=\Gamma_{\rm th}^{-1}$. Therefore,  $|\alpha_{\rm max}|\sim 1$ requires strong coupling conditions, which can be realistically achieved with  nanoresonators coupled to nitrogen-vacancy  centers in diamond~\cite{RablPRB2009,KolkowitzScience2012,WenzelProcRSocA2012} and have already been experimentally demonstrated with superconducting circuits~\cite{OConnellNature2010,PirkkalainenNature2013}.  Both approaches can be adapted to setups with fully levitated objects~\cite{RomeroIsartPRL2012,CirioPRL2012,Yin2013,Yin2013b}, where the mechanical trapping frequencies $\omega$ can be tuned close to zero and ultra-high Q-values are expected.

\emph{Quantum vs. classical correlations.} To show explicitly the difference between quantum and classical correlations, we repeat the above analysis with a qubit, which is modulated by an equivalent classical field $x_c(t)$ via the Hamiltonian $H_{\rm int}=\sqrt{2}\lambda x_c(t) |e\rangle\langle e|$. Since before each measurement the qubit is initialized in state $|g\rangle$, the value of $Z(t_n)$ only depends on the accumulated phase $\Phi_n =-\sqrt{2}\lambda \int_{t_n-\tau_n}^{t_n}   x_c(s) ds$, and for a fixed trajectory $x_c(t)$ we obtain $p_{\pm}(t_n)=(1\pm \cos(\varphi_n+\Phi_n))/2$. A general two time correlation function is then given by
\begin{equation}\label{eq:ZZclassical}
\begin{split}
\langle Z(t_i)Z(t_j)\rangle = \int & d\Phi_i d\Phi_j P(\Phi_i,\Phi_j)  \\
 &\times \cos(\varphi_i + \Phi_i)  \cos(\varphi_j + \Phi_j),
\end{split} 
\end{equation}
where $P(\Phi_i,\Phi_j)$ is the joint probability function for obtaining $\Phi_i$ and $\Phi_j$.  Since the $P(\Phi_i,\Phi_j)$ are derived from a single probability distribution for the underlying process $x_c(t)$ (which is independent of the measurement), the correlations in Eq.~\eqref{eq:ZZclassical} are necessarily bound by the LGI~\cite{Nori,SuppInfo}. As a specific example we consider the field  $ x_c(t)=A \cos(\omega t +\delta_0)$, with a random initial phase $\delta_0$ and a thermal distribution of the amplitude $A$. 
In this case we obtain the correlation function~\cite{SuppInfo}
 \begin{equation}
 \begin{split}
C(t_1,t_2) = \frac{1}{2}&\Big[ \cos(\varphi_1+\varphi_2) e^{- 2|\alpha(\tau)|^2 \langle x_c^2\rangle(1+Ê\cos \theta)}\\
 +&  \cos(\varphi_1-\varphi_2) e^{- 2 |\alpha(\tau)|^2 \langle x_c^2\rangle (1-Ê\cos \theta)} \Big].
 \end{split}
 \end{equation} 
This result closely resembles the quantum mechanical result in Eq.~\eqref{eq:Cij}, apart from the phase  $\gamma=|\alpha|^2\sin \theta$. In the quantum mechanical model this phase arises from the non-commutativity of the two displacement operations  
and is responsible for the violation of the LGI.

\emph{Conclusions and outlook.} In summary, we described an experimentally feasible scheme for probing macroscopic quantum superpositions via Ramsey correlation measurements. We  explicitly discussed the application of this technique for testing LGIs for massive mechanical resonators, but the method could be applied for other tests of quantum mechanics as well. For example, in the special case of commuting displacement operators the three-point correlations given in Eq.~\eqref{eq:Correlations} reduce to
\begin{equation}\label{eq:Joint}
\langle Z(t_3)Z(t_2)Z(t_1)\rangle= {\rm Tr}\{Q_{t_3} Q_{t_2} Q_{t_1} \rho_0 \},
\end{equation} 
and measures correlation functions of three modular variables. Plastino and Cabello~\cite{PlastinoPRA2010} have shown that such correlation functions can be used to test quantum contextuality in a two mode continuous variable system and a generalization of our method to two or multiple modes~\cite{WorkInProgress} provides a simple technique to implement such and related experiments.
It must be acknowledged that assumptions (i) and (ii) can only be tested jointly, and therefore the non-invasive measurability is assumed here to be valid in a macrorealistic theory~\cite{Leggett}.
In addition, specific invasive hidden variable models can be excluded by complementary measurements~\cite{WildeFoundPhys2012} and for more rigorous studies a generalization of our scheme to space-like separated resonators can be envisioned, where any invasive influence of measurements must be superluminal in order to explain quantum correlations.

\emph{Acknowledgments.} The authors thank S. Bennett, M. Lukin and M. Aspelmeyer for stimulating discussion.    
This work was supported by the John Templeton Foundation and the Austrian Science Fund (FWF) through SFB FOQUS, Individual project 2462 and the START grant Y 591-N16.

\widetext

\begin{center}
{\bf SUPPLEMENTARY MATERIAL}
\end{center}
\vspace{0.5cm}

This supporting information provides detailed derivations of various results presented in the main part of the paper and generalizes the analysis to modulated qubit-resonator couplings for amplifying the displacement amplitude.  

\section{\textbf{I}. State dependent displacements $\&$ modulated couplings} \label{sec:DrivenHarmonicOscillator} 
We first present the general results for the evolution operator generated by Hamiltonian (1) in the paper. Compared to the purely static coupling discussed in the main text, here we also include the possibility to modulate the coupling by applying fast $\pi$-pulses to flip the state of the qubit during the measurement process~\cite{TianPRA2005,BennettNJP2012}. These spin flips are taken into account by changing into a `toggling frame'~\cite{BennettNJP2012}, where the total Hamiltonian (in the frame rotating with the bare qubit frequency $\omega_q$) is given by
\begin{equation}\label{eq:Htotal}
H(t)=  \omega_m a^\dag a   +\left[\Delta(t)+ \lambda (a+a^\dag)\right]\left(f_g(t)|g\rangle\langle g| +  f_e(t) |e\rangle\langle e|\right).
\end{equation}
The function $f_e(t)=1,0$, where $f_e(t=0)=1$, tracks the qubit population, which is initially in the excited state and changes between $0$ and $1$ every time the qubit is flipped. Similarly, $f_g(t)=1-f_e(t)$. For completeness, Eq.~\eqref{eq:Htotal} also includes a classical frequency shift $\Delta(t)$.  We write the total evolution operator generated by $H(t)$ as
\begin{equation}\label{eq:Ut1}
U(t) = \mathcal{T} e^{-i \int_0^t H(s)ds}=  U_0(t) \left[ \tilde U_g(t) |g\rangle\langle g| + \tilde U_e(t)  |e\rangle\langle e|  \right], 
\end{equation}
where $U_0(t)=e^{-i\omega  a^\dag a t} $ is the free evolution of the resonator and 
\begin{equation}\label{eq:dotUtilde}
\partial_t \tilde U_{e,g}(t)= -i   f_{e,g}(t) \left[  \Delta(t)+ \lambda\left(a e^{-i\omega t}+a^\dag e^{i\omega t} \right) \right] \tilde U_{e,g}(t).
\end{equation} 
This equation can be solved by the ansatz
\begin{equation}
\tilde U_{e,g}(t)= e^{i\phi_{e,g}(t)}  \mathcal{D}(\tilde \alpha_{e,g}(t)),
\end{equation}
where $\mathcal{D}(\alpha)=e^{\alpha a^\dag- \alpha^*a}$ is the displacement operator.  By reinserting this ansatz back into Eq.~\eqref{eq:dotUtilde} we obtain
\begin{equation}
\dot{\tilde \alpha}_{e,g}= -i\lambda f_{e,g}(t)  e^{i\omega t}, \qquad \dot \phi_{e,g}= \frac{i}{2}\left( \tilde \alpha_{e,g} \dot{\tilde \alpha}^*_{e,g} -  \dot{\tilde \alpha}_{e,g} \tilde \alpha_{e,g}^*\right)-  f_{e,g}(t) \Delta(t),
\end{equation} 
with general solutions 
\begin{equation}
\tilde{\alpha}_{e,g}(t)=-i\lambda \int_0^t ds\,  f_{e,g}(s)  e^{i\omega s},  
\end{equation}  
and
\begin{equation}
\phi_{e,g}(t)=\lambda^2 \int_0^t ds_1 \int_0^{s_1} ds_2\, f_{e,g}(s_1)f_{e,g}(s_2)\sin(\omega(s_1-s_2)) -  \int_0^t ds \, f_{e,g}(s)\Delta(s).
\end{equation}
For the analysis of the correlation functions discussed below  we revert the ordering between the free evolution and the displacement operators in Eq.~\eqref{eq:Ut1}, and write the evolution operator as
\begin{equation}
U(t)=  U_\alpha(t) U_0(t),
\end{equation}
where 
\begin{equation}
U_\alpha(t)=   \left[ e^{i\phi_{g}(t)}  \mathcal{D}( \alpha_{g}(t)) |g\rangle\langle g| + e^{i\phi_{e}(t)}  \mathcal{D}(\alpha_{e}(t))  |e\rangle\langle e|  \right].
\end{equation}
In this case the displacement amplitudes
\begin{equation}
\alpha_{e,g}(t)= e^{-i\omega t} \tilde \alpha_{e,g}(t)=  -i\lambda \int_0^t ds\,  f_{e,g}(s)  e^{-i\omega (t- s)} ,
\end{equation} 
correspond to the coherent state amplitudes generated from a resonator initially prepared  in the ground state. 

\subsection{\textbf{A}. Static coupling}
For a static coupling we have $f_e(t)=1$ and $f_g(t)=0$. Therefore, $\alpha_g(t)=\phi_g(t)=0$ and 
\begin{equation}
\alpha(t)\equiv\alpha_{e}(t)= \frac{\lambda}{\omega}\left(e^{-i\omega t}-1\right),\qquad \phi(t)\equiv \phi_e(t)= \frac{\lambda^2}{\omega^2}\left(\omega t- \sin \omega t \right) - \phi_c(t).
\end{equation}
These are the results discussed in the main part of the paper, with an additional phase $\phi_c(t)=  \int_0^t ds \, \Delta(s)$ from classical modulations of the qubit frequency.

\subsection{\textbf{B}. Modulated coupling}
The maximal state dependent displacement amplitude can be amplified by periodically flipping the qubit state during the measurement. In this case both qubit components get displaced, $\alpha_{e,g}(t)\neq 0$, according to the results derived above. In Sec. \textbf{II} below we show that for the creation and detection of superposition states or for the measurement of single modular variables the relevant quantities are the relative displacement amplitude 
\begin{equation}\label{eq:alphatot}
\alpha(t)\equiv \alpha_e(t)-\alpha_g(t)= - i\lambda \int_0^t ds \, f_-(s) e^{-i\omega(t- s)},
\end{equation} 
and the total phase
\begin{equation}\label{eq:phitot}
\begin{split}
\phi(\tau)\equiv&\phi_e(\tau)-\phi_g(\tau)-{\rm Im}\{\alpha_g(\tau) \alpha_e^*(\tau)\} 
=  \lambda^2 \int_0^t dt_1 \int_0^{t_1} dt_2\, f_-(t_1) f_+(t_2)\sin(\omega(t_1-t_2)) -  \int_0^t ds \, f_-(s)\Delta(s),
\end{split}
\end{equation}
where we have defined
\begin{equation}
 f_\pm(t)=f_e(t)\pm f_g(t).
\end{equation} 
For a total number of $N_p$ equally spaced $\pi$-pulses separated by a time $\tau_p$ we obtain
\begin{equation}
\begin{split}
\alpha(t)=& - i\lambda e^{-i\omega t} \sum_{n=0}^{N_p} (-1)^n  e^{i\omega n\tau_p} \int_0^{\tau_p}ds \, e^{i\omega s} = \frac{\lambda}{\omega}e^{-i\omega \tau_p N_p} \left(1-e^{i\omega \tau_p}\right)  \sum_{n=0}^{N_p}  \left(- e^{i\omega \tau_p}\right)^n. 
\end{split}
\end{equation}
For a resonant modulation, $\tau_p=\pi/\omega$, this leads to
\begin{equation}
\alpha\left(t=(N_p+1)\tau_p\right)=  (-1)^{N_p} \frac{2\lambda}{\omega} N_p,
\end{equation} 
and the displacement amplitude increases proportional to the number of applied pulses.

\subsection{C. Asymmetric displacement}

As we show in Sec. \textbf{II. D.} below,  for the violation of the Leggett-Garg inequality not only the relative amplitude $\alpha=\alpha_e-\alpha_g$ matters, but also the combination  $(\alpha_e-\alpha_g)(\alpha_e+\alpha_g)$ appears in the expression of the additional phase $\gamma$. The sum of the two displacements is given by
\begin{equation}
\alpha_e+\alpha_g= - i\lambda \int_0^t ds \, (f_e(s)+f_g(s)) e^{-i\omega(t- s)}=- i\lambda \int_0^t ds \, e^{-i\omega(t- s)}= \frac{\lambda}{\omega}\left(e^{-i\omega t}-1\right),
\end{equation}
and for a static coupling with a two level system it cannot be amplified by a simple $\pi$-pulse sequence. However, an asymmetric amplification $|\alpha_e|\gg|\alpha_g|$  can still be achieved in certain variations of the present system. One possibility is to implement directly a time dependent coupling $\lambda\rightarrow \lambda(t)$. This can be achieved, for a example, by coupling a resonator to a superconducting qubit and by modulating the charge on the resonator. Another possibility is to use a three level system as in the case of a NV center~\cite{RablPRB2009} with three spin states $|m_s=0,\pm1\rangle$. In this case $f_{e,g}(t)=0,\pm1$ and the following strategy can be applied: After creating the initial superposition between $|g\rangle=|m_s=0\rangle$ and $|e\rangle=|m_s=+1\rangle$ the state $|g\rangle$ is transferred to $|m_s=-1\rangle$, where is remains for the rest of the sequence.  Therefore, $f_g(t)=-1$ and if the total time $\tau$ is a multiple of $2\pi/\omega$, $\alpha_g(\tau)=0$. In turn, the state $|e\rangle$ can be flipped with  a period $\tau_p=\pi/\omega$ between $|m_s=1\rangle$ and $|m_s=0\rangle$ to implement a resonant amplification as discussed above.

\section{II. Ramsey measurements $\&$ correlation functions}\label{sec:Correlations}
In this section we detail the derivation of the correlation function for two and multiple Ramsey measurements (RMs). The following results are derived for a purely unitary evolution. The effect of dissipation is discussed in Sec. \textbf{IV}.   

\subsection{A. Ramsey measurements and modular variables} 
We first consider a single measurement. At the beginning of the Ramsey sequence the qubit is initialized in state $|g\rangle$. The pulse sequence specified in the main part of the paper corresponds to the evolution operator  
\begin{equation}
U_M(\varphi,\tau) = R_{\pi/2}(0) U_\alpha(\tau) U_0(\tau) R_{\pi/2}(\varphi),
\end{equation} 
acting on the combined qubit-resonator system. Here the $R_{\pi/2}(\varphi)$ denote $\pi/2$-rotations of the qubit with an adjustable phase $\varphi$, which in the basis $\{ |g\rangle, |e\rangle\}$ is  defined as 
\begin{equation}
R_{\pi/2}(\varphi)=\frac{1}{\sqrt{2}}\left(\begin{array}{cc}  1 & e^{i\varphi} \\  -e^{-i\varphi} & 1\end{array}\right).
\end{equation}
The evolution between the pulses, $U_\alpha(\tau) U_0(\tau)$, describes the qubit-resonator interaction for a measurement time $\tau$ as defined in Sec. \textbf{I}. If we assume that right before the measurement the resonator is prepared in a pure state $|\psi\rangle$ the state of the system after this pulse sequence is 
\begin{align}
U_{M}(\varphi,\tau)|g\rangle |\psi\rangle = &   E_-(\varphi,\tau) |\psi\rangle |g\rangle +   E_+(\varphi,\tau)|\psi\rangle | e\rangle, 
\end{align}
where 
\begin{equation}
E_\pm(\varphi,\tau)=  \frac{1}{2}\left(e^{i\phi_g(\tau)} \mathcal{D}(\alpha_g(\tau)) \pm e^{i (\phi_e(\tau)+\varphi) }  \mathcal{D}(\alpha_e(\tau)) \right)U_0(\tau),
\end{equation} 
are Kraus operators which act on the resonator state and fulfill the normalization condition
\begin{equation}
E_+^\dagger E_+ +E_-^\dagger E_-=\mathbbm{1}.
\end{equation}
More generally, if at the initial time $t_0$ the resonator is in an arbitrary state $\rho_0$, the total system density operator after the pulse sequence is
\begin{align}
U_M(\varphi,\tau)\left(|g\rangle\langle g| \otimes\rho_0\right) U^\dagger_M(\varphi,\tau)=|g\rangle\langle g| \otimes E_-(\varphi,\tau)\rho_0 E^\dagger_-(\varphi,\tau)+|e\rangle\langle e|\otimes E_+(\varphi,\tau)\rho_0 E^\dagger_+(\varphi,\tau) \\+|g\rangle\langle e|\otimes E_-(\varphi,\tau)\rho_0 E^\dagger_+(\varphi,\tau)+|e\rangle\langle g|\otimes E_+(\varphi,\tau)\rho_0 E^\dagger_-(\varphi,\tau).
\end{align}
Tracing over the spin degrees of freedom gives
\begin{equation}
\rho(t_1)=E_-(\varphi,\tau)\rho_0 E^\dagger_-(\varphi,\tau) + E_+(\varphi,\tau)\rho_0 E^\dagger_+(\varphi,\tau),
\end{equation}
for the \emph{unconditioned} resonator state at time $t_1=t_0+\tau$. The probabilities $p_+$ and $p_-$ for finding the qubit in state $|e\rangle$ and $|g\rangle$, respectively, are then given by 
\begin{equation} 
p_\pm(t_1) = {\rm Tr}\{ E_\pm(\varphi,\tau)\rho_0 E^\dagger_\pm(\varphi,\tau)\}.
\end{equation} 
Depending on the measurement outcome the \emph{conditioned} resonator state is
\begin{equation} 
\rho^\pm(t_1) =   \frac{  E_\pm(\varphi,\tau)\rho_0 E^\dagger_\pm(\varphi,\tau)}{p_\pm(t_1)}.
\end{equation} 
To simplify the expressions for the probabilities $p_\pm$ we use $\mathcal{D}(\alpha)\mathcal{D}(\beta)= e^{i{\rm Im}(\alpha \beta^* )} \mathcal{D}(\alpha+\beta)$ and write 
\begin{equation}
e^{i(\phi_e-\phi_g+\varphi)}   \mathcal{D}^\dag(\alpha_g(\tau)) \mathcal{D}(\alpha_e(\tau)) \equiv e^{i (\varphi+ \phi(\tau)) } \mathcal{D}(\alpha(\tau)),
\end{equation} 
where the amplitude difference $\alpha(\tau)=\alpha_e(\tau)-\alpha_g(\tau)$  and the total phase $\phi(\tau)$ are defined in Eqs.~\eqref{eq:alphatot} and \eqref{eq:phitot}, respectively.
By setting $\bar \varphi = \varphi + \phi(\tau)$ we obtain
\begin{equation}
p_\pm(t_1)= \frac{1}{2}\left[1\pm {\rm Tr}\{ÊÊÊQ(\bar \varphi, \alpha(\tau)) U_0(t_1)\rho_0U^\dag_0(t_1)\}\right], 
\end{equation} 
where  
\begin{equation}
Q(\varphi,\alpha) = \frac{1}{2}Ê\left( e^{i \varphi } \mathcal{D}(\alpha)+ e^{-i \varphi} \mathcal{D}^\dag(\alpha)\right) = \cos\left( \varphi+ i \alpha^* a - i \alpha a^\dag \right).
\end{equation}
If we denote by $Z(t_1)=\pm 1$ the dichotomic variable describing the measurement outcome at time $t_1=t_0+\tau$, we obtain 
\begin{equation}
\langle Z(t_1)\rangle = p_+(t_1)-p_-(t_1)= \rm Tr\{Q(\bar \varphi,\alpha_1)\rho_0\},
\end{equation} 
where we have used $U^\dag_0(t_1)Q(\bar \varphi, \alpha(\tau))U_0(t_1)=Q(\bar \varphi,\alpha_1)$ with the convention $\alpha_n=e^{i\omega t_n} \alpha(\tau_n)$. Therefore, a single RM measures the expectation value of the modular variable $Q(\bar \varphi,\alpha_1)$ with respect to $\rho_0$.  By writing $\alpha_1=\alpha_R+i\alpha_I$ we finally obtain
\begin{equation}
Q(\bar \varphi,\alpha_1) =  \cos\left( \bar \varphi + \sqrt{2}\alpha_I \, x - \sqrt{2} \alpha_R \,p  \right),
\end{equation}
where $x=(a+a^\dag)/\sqrt{2}$ and $p=i(a^\dag-a)/\sqrt{2}$.

\subsection{B. Correlations: general results}Ê
We now generalize the above considerations for a sequence of measurements. We set $t_0=0$ the time right before the first measurement and denote by $t_n$ the time after the $n$-th RM is complete. The variables $Z(t_n)=\pm 1$ describe the outcome of the respective measurements. Each Ramsey sequence is characterized by displacement amplitudes $\alpha_e(\tau_n)$ and $\alpha_g(\tau_n)$, an adjustable phase of the first $\pi/2$-pulse $\varphi_n$ and the geometric phases $\phi_{e,n}\equiv\phi_{e,n}(\tau_n)$ and $\phi_{g,n}\equiv\phi_{g,n}(\tau_n)$ as defined above.  Starting from the initial resonator density operator $\rho_0=\rho(t_0)$ the state condition on the first measurement outcome $\eta_1=\pm$ is
\begin{equation}
\rho^{\eta_1}(t_1)=E_{\eta_1}(\tau_1) \rho_0 E_{\eta_1}^\dag(\tau_1)/p_{\eta_1}. 
\end{equation} 
This state evolves freely for a time $t_2-\tau_2- t_1$ and then a second measurement is performed. By repeating the arguments from above, the probabilities for this second measurement, conditioned on the first outcome, are given by  
\begin{equation}
p_{\pm|\eta_1}= \frac{1}{2}\left(1\pm {\rm Tr}\{ÊÊÊQ(\bar \varphi_2,\alpha(\tau_2))U_0(t_2-t_1) \rho^{\eta_1}(t_1) U_0^\dag(t_2-t_1)  \}\right).
\end{equation} 
Therefore, the conditioned expectation value of the second measurement is
\begin{equation}
\langle Z(t_2)\rangle_{\eta_1}=p_{+|\eta_1}-p_{-|\eta_1} = {\rm Tr}\{ÊÊÊQ(\bar \varphi_2,\alpha_2) \rho^{\eta_1}(t_2) \},
\end{equation}
where $\rho^{\eta_1}(t_2)=U_0(t_2-t_1) \rho_1^{\eta_1}(t_1) U_0^\dag(t_2-t_1) $ denotes the time evolved conditioned density operator.
The two point correlation function between two successive measurements is 
\begin{equation}
\begin{split}
\langle Z(t_2)Z(t_1)\rangle &= (p_{+|+}-p_{-|+})p_++ (p_{-|-}-p_{+|-})p_-\\
&={\rm Tr}\{ÊÊÊQ(\bar \varphi_2,\alpha_2) ( p_+ \rho^{+}(t_2)- p_-\rho^{-}(t_2)) \}. 
\end{split}
\end{equation}
To write the resulting expression in a compact form  we define the following superoperator 
\begin{equation}Ê
\mathcal{Q}_{t_n} \rho :=  \frac{1}{2} \left( e^{i(\phi_{e,n}+\varphi_n-\phi_{g,n})} \mathcal{D}(\alpha_{e,n}) \rho  \mathcal{D}^\dag(\alpha_{g,n}) +e^{-i(\phi_{e,n}+\varphi_n-\phi_{g,n})} \mathcal{D}(\alpha_{g,n}) \rho  \mathcal{D}^\dag(\alpha_{e,n})  \right),
\end{equation} 
where as above we have used the abbreviation $\alpha_{e,n}= e^{i\omega (t_n-t_0)}\alpha_e(\tau_n)$, etc. In the following it is assumed $t_0=0$ for brevity.
With this notation we obtain
\begin{equation}
\langle Z(t_1)\rangle = {\rm Tr}\{Ê\mathcal{Q}_{t_1} \rho_0\},
\end{equation}
and 
\begin{equation}
\langle Z(t_2)Z(t_1)\rangle = {\rm Tr}\{Ê\mathcal{Q}_{t_2} \mathcal{Q}_{t_1} \rho_0\}.
\end{equation}
The same analysis can be generalized to higher order correlation functions. For example,
\begin{equation}
\begin{split}
\langle Z(t_3)Z(t_2)Z(t_1)\rangle = & p_+(t_1)\langle Z(t_3)Z(t_2)\rangle_{\eta_1=+}- p_-(t_1)\langle Z(t_3)Z(t_2)\rangle_{\eta_1=-}= {\rm Tr}\{\mathcal{Q}_{t_3}Ê\mathcal{Q}_{t_2} \mathcal{Q}_{t_1} \rho_0\},
\end{split} 
\end{equation} 
where $\langle \cdot\rangle_{\eta_1}$ denotes the average with respect to the conditioned density operator after the first measurement. By iterating this argument
we obtain
\begin{equation}
\langle Z(t_n) \dots Z(t_2) Z(t_1)\rangle =  {\rm Tr}\{\mathcal{Q}_{t_n} \dots Ê\mathcal{Q}_{t_2} \mathcal{Q}_{t_1} \rho_0\}.
\end{equation}

\subsection{C. Two-time correlations: static coupling} 
We now evaluate the two time correlation function $\langle Z(t_2)Z(t_1)\rangle$ for the case of static coupling discussed in the main part of the paper. In this case $\alpha_{g}=\phi_g=0$ and  we obtain  
\begin{equation}Ê
\mathcal{Q}_{t_n} \rho :=  \frac{1}{2} \left( e^{i\bar \varphi_n} \mathcal{D}(\alpha_{n}) \rho  +e^{-i\bar \varphi_n}  \rho  \mathcal{D}^\dag(\alpha_{n})  \right),
\end{equation} 
where $\bar \varphi_n= \varphi_n+\phi_{e,n}$, $\alpha_n\equiv \alpha_{e,n}=  \alpha(\tau_n) e^{i\omega t_n}$.
The full expression for the correlation function is
\begin{equation}
\begin{split}
\langle Z(t_2)Z(t_1)\rangle=&\frac{1}{4}{\rm Tr}\left\{\left(e^{i\bar \varphi_2} \mathcal{D}(\alpha_{2})  +  e^{-i\bar \varphi_2} \mathcal{D}^\dag(\alpha_{2})\right)  \left( e^{i\bar \varphi_1} \mathcal{D}(\alpha_{1})\rho_0 + \rho_0 e^{-i\bar \varphi_1}  \mathcal{D}^\dag(\alpha_{1})    \right)   \right\}\\
=&\frac{1}{4}\left[ \langle e^{i(\bar \varphi_1+\bar \varphi_2)} \mathcal{D}(\alpha_{2})\mathcal{D}(\alpha_{1})\rangle + \langle e^{i(\bar \varphi_1- \bar \varphi_2)} \mathcal{D}^\dag(\alpha_{2})\mathcal{D}(\alpha_{1})\rangle   + {\rm c.c.} \right].
\end{split}
\end{equation} 
Using
\begin{equation}
\begin{split}
 \mathcal{D}(\alpha_{2})\mathcal{D}(\alpha_{1}) =& e^{i {\rm Im}\{\alpha_{2} \alpha^*_{1}  \}}  \mathcal{D}(\alpha_{1}+\alpha_{2}),\\ \mathcal{D}^\dag(\alpha_{2})\mathcal{D}(\alpha_{1}) = &e^{-i {\rm Im}\{Ê\alpha_{2} \alpha^*_{1}Ê  \}}  \mathcal{D}(\alpha_{1}-\alpha_{2}),
\end{split}
\end{equation}
we finally obtain
\begin{equation}
\begin{split}
\langle Z(t_2)Z(t_1)\rangle=&\frac{1}{2}\left[\cos(\bar \varphi_1+\bar \varphi_2+ \gamma)  \langle \mathcal{D}(\alpha_{1}+\alpha_{2})\rangle  +  \cos(\bar \varphi_1-\bar \varphi_2-\gamma)  \langle \mathcal{D}(\alpha_{1}-\alpha_{2})\rangle  \right],
\end{split}
\end{equation}
where
\begin{equation}
\gamma= {\rm Im}\{\alpha_{2} \alpha^*_{1}\} = {\rm Im}\{\alpha(\tau_2) \alpha^*(\tau_1) e^{i\omega(t_2-t_1)}\} .
\end{equation} 
For identical measurements $\alpha(\tau_i)=\alpha$, $\theta=\omega(t_2-t_1)$ and assuming that $\rho_0$ is in a thermal state with mean occupation $\bar n$ this result leads to Eq. (10) in the main part of the paper.

\subsection{D. Two-time correlations: modulated coupling}\label{subsec:CorrelationsModulated}
We now consider the case of a modulated coupling where $\alpha_{e,n}\neq0$ and $\alpha_{g,n}\neq 0$. As already shown above, for the second measurement only the difference $\alpha_2=\alpha_{e,2}-\alpha_{g,2}$ and the combined phase $\phi_2$ as defined in Eq.~\eqref{eq:phitot} appear in the expectation value, i.e.,  
\begin{equation}
{\rm Tr}\left\{ \mathcal{Q}_{t_2} \mathcal{Q}_{t_1}\rho_0  \right\} = \frac{1}{2}{\rm Tr}\left\{ \left(e^{i(\varphi_2+\phi_2)} \mathcal{D}(\alpha_{2})  +  e^{-i(\varphi_2+\phi_2)} \mathcal{D}^\dag(\alpha_{2})\right) \mathcal{Q}_{t_1}\rho_0  \right\}. 
\end{equation} 
By introducing again $\bar \varphi_2=\varphi_2+\phi_2$ the total correlation function is given by
\begin{equation}
\begin{split}
\langle Z(t_2)Z(t_1)\rangle=&\frac{1}{4}\left[Ê{\rm Tr}\left\{ \mathcal{D}^\dag(\alpha_{g,1})   \left(e^{i\bar \varphi_2} \mathcal{D}(\alpha_{2})  +  e^{-i\bar \varphi_2} \mathcal{D}^\dag(\alpha_{2})\right)   \mathcal{D}(\alpha_{e,1} )\rho_0   \right\} + {\rm c.c.}\right].
\end{split}
\end{equation} 
We reorder the operators according to 
\begin{equation}
\begin{split}
e^{i(\phi_{e,1}-\phi_{g,1}+\varphi_1)}  \mathcal{D}^\dag(\alpha_{g,1}  ) \mathcal{D}(\alpha_{2})  \mathcal{D}(\alpha_{e,1}  ) = e^{i(\phi_1+\varphi_1)}Ê  e^{i\phi^\prime} \mathcal{D}(\alpha_{2}) \mathcal{D}(\alpha_{1}), \\ 
e^{i(\phi_{e,1}-\phi_{g,1}+\varphi_1)}  \mathcal{D}^\dag(\alpha_{g,1} ) \mathcal{D}^\dag (\alpha_{2})  \mathcal{D}(\alpha_{e,1}  ) = e^{i(\phi_1+\varphi_1)}Ê  e^{-i\phi^\prime} \mathcal{D}^\dag(\alpha_{2}) \mathcal{D}(\alpha_{1}),
\end{split}
\end{equation}
where $\alpha_{1}= (\alpha_{e,1}-\alpha_{g,1})e^{i\omega t_1}$, $\phi_1$ is the combined phase [as defined in Eq.~\eqref{eq:phitot}] of the first measurement and
\begin{equation}
\phi^\prime = 2{\rm Im}\{Ê\alpha_{g,1}^*\alpha_2    \}.Ê
\end{equation} 
In total we obtain
\begin{equation}
\begin{split}
\langle Z(t_2)Z(t_1)\rangle=&\frac{1}{2}\left[\cos(\bar \varphi_1+\bar\varphi_2+ \gamma)  \langle \mathcal{D}(\alpha_{t_1}+\alpha_{t_2})\rangle  +  \cos(\bar\varphi_1-\bar\varphi_2-\gamma)  \langle \mathcal{D}(\alpha_{t_1}-\alpha_{t_2})\rangle  \right],
\end{split}
\end{equation}
where now
\begin{equation} 
\gamma =  {\rm Im}\{\alpha_{2} \alpha_{1}^* \} + 2{\rm Im}\{Ê\alpha_{g,1}^*\alpha_2  \} =  {\rm Im}\{\alpha(\tau_2) (\alpha_{e}(\tau_1)+\alpha_{g}(\tau_1))^* e^{i\omega(t_2-t_1)} \}. 
\end{equation} 
This means that the correlation functions for a modulated coupling are essentially the same as in the case for a static coupling, but with $\alpha=\alpha_{e}-\alpha_g$. However, there is a difference in the geometric phase $\gamma$, which scales as
\begin{equation}
\gamma \sim (\alpha_{e,2}-\alpha_{g,2}) (\alpha_{e,1}+\alpha_{g,1})^*.
\end{equation}
Therefore, for the violation of the Leggett-Garg inequality it is advantageous to have an asymmetric displacement $|\alpha_e|\gg1$ but $|\alpha_g|\ll |\alpha_e|$, which can be achieved with the methods outlined in the end of Sec. \textbf{I}.

\section{III. Classical correlations} 
In this section we now re-evaluate the Ramsey correlation measurement for a Hamiltonian
\begin{equation}
H= \sqrt{2} \lambda x_c(t) |e\rangle\langle e|,
\end{equation}     
which describe the case where the qubit frequency is modulated by a classical random field $x_c(t)$ (the factor $\sqrt{2}$ has been introduce to have the direct correspondence $x_c\leftrightarrow x=(a+a^\dag)/\sqrt{2}$). 
Since before each measurement the qubit is initialized in state $|g\rangle$ and has no memory of previous events, the measurement outcome at time $t_n$ is only a function of the accumulated phase
\begin{equation}
\Phi_n =-\sqrt{2}\lambda  \int_{t_n-\tau_n}^{t_n}   x_c(s) ds.
\end{equation}  
More precisely, the probabilities to measure $Z(t_n)=\pm1$ are given by 
\begin{equation}Ê
p_\pm(t_n)=\int d\Phi_n \, P_n(\Phi_n)  \frac{1}{2} \left(1\pm \cos(\varphi_n + \Phi_n)\right) , 
\end{equation}
where $P_n(\Phi_n)$ is the classical probability distribution for the accumulated phase $\Phi_n$. Similarly, the joint probabilities for two measurements are, for example,  
\begin{equation}\label{eq:pClassical}
p_{++}(t_n,t_m)=\int d\Phi_n d\Phi_m  \, P_{nm}(\Phi_n,\Phi_m)  \frac{1}{2} \left(1+  \cos(\varphi_n + \Phi_n)\right) \frac{1}{2} \left(1+  \cos(\varphi_n + \Phi_m)\right),
\end{equation}
where $P_{nm}(\Phi_n,\Phi_m)$ is the joint probability distribution for $\Phi_n$ and $\Phi_m$, and so on. A general two point correlation is then given by
\begin{equation}\label{eq:ZZclassicalSupp}
\langle Z(t_n)Z(t_m)\rangle = \int d\Phi_n d\Phi_m  P_{nm}(\Phi_n,\Phi_m)  \cos(\varphi_n + \Phi_n)  \cos(\varphi_m + \Phi_m).
\end{equation}
For the derivation of the Wigner-type LGI it is assumed that the $P_n(\Phi_n)$ and $P_{nm}(\Phi_n,\Phi_m)$ are marginals of a \emph{single} joint distribution $P(\Phi_1,\Phi_2,\Phi_3)$~\cite{Nori}, e.g., 
\begin{equation} 
P_{13}(\Phi_1,\Phi_3)\equiv P(\Phi_1,\Phi_3) =\int d\Phi_2   \, P(\Phi_1,\Phi_2,\Phi_3).
\end{equation}
This conditions follows from a macrorealistic description of $x_c(t)$ and the assumptions of non-invasive measurements. ÊIf this property holds for the $P(\Phi_n,\Phi_m)$ it is also true for the probabilities $p_{\eta_n\eta_m}(t_n,t_m)$, e.g., 
\begin{equation}
p_{++}(t_1,t_3) = \sum_{\eta_2=\pm}   p_{+\eta_2+}(t_1,t_2,t_3),
\end{equation} 
etc. Therefore, under these conditions the proof for the LGI holds for the correlations $\langle Z(t_n)Z(t_m)\rangle$. 

\subsection{A. Example: classical thermal oscillator} 
To illustrate the difference between quantum and classical correlation we now consider the coupling of the measurement qubit to a classical oscillating field of the form 
\begin{equation}
x_c(t)= A \cos(\omega t +\delta_0).
\end{equation}  
For a fixed amplitude $A>0$ and initial phase $\delta_0$ we obtain
\begin{equation}
\Phi_n= A\frac{\sqrt{2}\lambda}{\omega} \left[ \sin(\omega t_n-\omega\tau_n+\delta_0)- \sin(t_n\omega+\delta_0)  \right] =- \frac{\sqrt{8}A\lambda}{\omega}\cos( (t_n+\tau_n/2)\omega+\delta_0)\sin(\tau_n\omega/2).
\end{equation} 
For comparison with the quantum mechanical correlations discussed in the paper we consider the following random distribution 
\begin{equation}
P(A,\delta_0)= \frac{A} {2\pi \langle x_c^2\rangle}  e^{- \frac{A^2}{2\langle x_c^2\rangle}},
\end{equation} 
which corresponds to that of an oscillating field in a thermal state with variance $\langle x_c^2\rangle$. For the first measurement we obtain
 \begin{equation}
 \langle Z(t_1)\rangle = \int_0^{2\pi}  d\delta_0  \int_0^\infty dA \, P(A,\delta_0)  \cos\left(\varphi_1- \frac{\sqrt{8}A\lambda}{\omega}\cos( (t_1+\tau_1/2)\omega+\delta_0)\sin(\tau_1\omega/2) \right).
 \end{equation}    
Using the Jacobi-Anger identity 
\begin{equation}
\cos(\varphi+a\cos(\delta))= \frac{1}{2}Ê\sum_{n=-\infty}^\infty \left[i^n e^{i\varphi} +(-i)^n e^{-i\varphi}\right] J_n(a) e^{i\delta n },
\end{equation} 
and the integrals 
\begin{equation}
\frac{1}{2\pi} \int_0^{2\pi}d\delta\,\cos(\varphi+a\cos(\delta)) = \cos(\varphi) J_0(a),\qquad  \frac{1}{a^2} \int_0^\infty dx  \, x e^{- x^2/(2a^2)}J_0(x b )=e^{-a^2b^2/2},
\end{equation} 
we obtain
\begin{equation}
 \langle Z(t_1)\rangle= \cos(\varphi_1) e^{- \frac{4\lambda^2\langle x_c^2\rangle}{\omega^2}\sin^2(\frac{\omega\tau_1}{2})}= \cos(\varphi_1) e^{- |\alpha(\tau_1)|^2\langle x_c^2\rangle},
\end{equation}
where we have used $\alpha(\tau)=\lambda/\omega(e^{-i\omega\tau}-1)$ for a static coupling. 
This expectation value exhibits similar collapse and revivals signatures as obtained from the quantum mechanical calculation. 
%
For the two-point correlation function with $\tau_1=\tau_2=\tau$ we obtain
 \begin{equation}
 \begin{split}
 \langle Z(t_2) Z(t_1)\rangle = \int_0^{2\pi}  d\delta_0  \int_0^\infty dA \, P(A,\delta_0) & \cos\left(\varphi_1- \frac{\sqrt{8}A\lambda}{\omega}\cos( (t_1+\tau/2)\omega+\delta_0)\sin(\tau\omega/2) \right)\\
 & \times \cos\left(\varphi_2- \frac{\sqrt{8}A\lambda}{\omega}\cos( (t_2+\tau/2)\omega+\delta_0)\sin(\tau\omega/2) \right).
 \end{split}
 \end{equation}    
Similar as above, we first integrate over the uniform $\delta_0$ distribution, where we use
\begin{equation}
\begin{split}
&\frac{1}{2\pi}Ê\int_0^{2\pi} d\delta\, \cos(\varphi_1+a_1\cos(\delta))\cos(\varphi_2+a_2\cos(\delta+\theta))\\
&= \frac{1}{4}Ê\left[   e^{i(\varphi_1+\varphi_2)} J_0\left( |a_1+a_2e^{i\theta}|\right) + e^{i(\varphi_1-\varphi_2)} J_0\left( |a_1-a_2e^{i\theta}|\right) + {\rm c.c.} \right].
\end{split}
\end{equation} 
Finally, after performing the integral over $A$ and setting $\theta=\omega(t_2-t_1)$ we end up with
 \begin{equation}
 \begin{split}
 \langle Z(t_2)Z(t_1)\rangle = \frac{1}{2}\left[ \cos(\varphi_1+\varphi_2) e^{- 2 |\alpha(\tau)|^2\langle x_c^2\rangle\left[1+Ê\cos(\theta)\right]}+  \cos(\varphi_1-\varphi_2) e^{- 2 |\alpha(\tau)|^2 \langle x_c^2\rangle\left[1-Ê\cos(\theta)\right]} \right].Ê
 \end{split}
 \end{equation} 
Again, the general form of this classical correlation function is very similar to the quantum mechanical predictions. However, in the quantum case there appears an additional phase $\cos(\varphi_1\pm \varphi_2) \rightarrow \cos(\varphi_1\pm\varphi_2 \pm |\alpha(\tau)|^2\sin( \theta))$, which is responsible for the appearance of non-classical correlations.

\section{IV. Decoherence}\label{sec:Decoherence} 

In this section we evaluate the effect of qubit decoherence and mechanical decoherence due to a weak coupling of the resonator to a finite temperature bath. We model these decoherence processes by a master equation of the form
\begin{equation}\label{eq:MasterEq}
\dot \rho =\mathcal{L}\rho= -i [H,\rho] +\mathcal{L}_q \rho + \mathcal{L}_m \rho,
\end{equation}
where 
\begin{equation}
\mathcal{L}_q \rho =\frac{1}{2T_2} \left( \sigma_zÊ\rho \sigma_z-\rho\right),
\end{equation} 
describes the qubit dephasing with dephasing time $T_2$ and 
\begin{equation}
\mathcal{L}_m \rho= \frac{\Gamma}{2} (N+1)(2a\rho a^\dag -a^\dag a \rho  - \rho a^\dag a ) + \frac{\Gamma}{2} N (2a^\dag \rho a -a a^\dag \rho  - \rho a a^\dag ),
\end{equation} 
describes the mechanical dissipation, where $\Gamma=\omega/Q$ is the mechanical damping rate for mechanical resonator with quality factor $Q$ and $N=1/(e^{\hbar\omega/k_BT}-1)$ is the equilibrium occupation number. In the high temperature limit $N\gg1$ we obtain $\Gamma_{\rm th}=\Gamma N\simeq k_BT/(\hbar Q)$ as the relevant mechanical decoherence rate.

\subsection{A. Decay of correlations} 
We first consider the limit $\Delta t \gg \tau$, where the effect of decoherence during the Ramsey sequences can be neglected and we denote by $\rho^{\eta_1=\pm}(t_1)$ the conditioned density operator after the first measurement. In this case the conditioned probabilities for the second measurement are 
\begin{equation}Ê
p_{\pm |\eta_1}(t_2)= \frac{1}{2}Ê\left(1\pm  {\rm Re}\{ e^{i\bar \varphi_2} \chi_{\eta_1}(\alpha(\tau_2), t_2)\}\right),
\end{equation}
where $\chi_{\eta_1}(\beta, t_2)=\text{Tr}\{\mathcal{D}(\beta)\rho^{\eta_1}(t_2)\}$ is the characteristic function of the density operator 
 $\rho^{\eta_1}(t_2)=e^{\mathcal{L}(t_2-t_1)} \rho^{\eta_1}(t_1)$. 
From the master equation~\eqref{eq:MasterEq} we can derive  the Fokker-Planck equation
\begin{equation}
\label{Eqcharact}
\dot \chi_{\eta_1}(\beta) = \left[i\Omega \beta \frac{\di}{\di \beta} -i\Omega^*\beta^* \frac{\di}{\di \beta^*}- \frac{\Gamma}{2}(2N+1) |\beta|^2 \right]\chi_{\eta_1}(\beta),
\end{equation}
for the evolution of $\chi_{\eta_1}(\beta, t)$, where $\Omega=\omega+i\Gamma/2$. The general solution of this equation is
\begin{equation}
\begin{split}
\chi_{\eta_1}(\beta, t_2)= e^{-(N+1/2)|\beta|^2(1-e^{-\Gamma (t_2-t_1)})} \chi_{\eta_1}\left(\beta e^{(i\omega-\Gamma/2)(t_2-t_1)},t_2=t_1\right),
\end{split}Ê
\end{equation}
from which we obtain the evolution of the conditioned probabilities $p_{\pm |\eta_1}(t_2)$ and the two-time correlation function $C(t_1,t_2)\equiv C(\alpha_2, \Delta t\simeq t_2-t_1)$, 
\begin{equation}
C (\alpha_2, \Delta t) = e^{-(N+1/2)|\alpha_2|^2(1-e^{-\Gamma \Delta t})} C(\alpha_2 e^{-\frac{\Gamma \Delta t}{2}}, \Delta t)\Big|_{\Gamma=0}.
\end{equation}
The Wigner function of the initial superposition state $|\psi^+\rangle$ plotted in Fig. 2Êa) of the main text is defined as
\begin{align}
\nonumber
&\mathcal{W}(\xi,\xi^*)= \dfrac{1}{\pi^2}\int d^2\beta  \,e^{-\beta\xi^*+\beta^*\xi} \chi_+(\beta,t_1).
\end{align}
During the waiting time $\Delta t$ it evolves as \cite{CatFinitTemp}
\begin{align}
\nonumber
 \mathcal{W}(\xi,\xi^*,\Delta t)=&\dfrac{1}{2 \pi \nu  p_+}  \Big[e^{\frac{-2\abs{\xi}^2}{\nu}}+ e^{\frac{-2\abs{\xi-\alpha_1 e^{-\Gamma \Delta t/2}}^2}{\nu }} \\ \nonumber
&+ 2 e^{\frac{-2\abs{\xi-\frac{\alpha_1}{2} e^{-\Gamma \Delta t/2}}^2}{\nu}}  e^{-\frac{\abs{\alpha_1}^2}{2}(1-\frac{e^{-\Gamma \Delta t}}{\nu})}   \cos\left(\varphi_1+\frac{2}{\nu}{\rm Im} \xi^* \alpha_1e^{-\frac{\Gamma \Delta t}{2}}\right)\Big],
\end{align}
where we defined $\nu=1+2N(1-e^{-\Gamma \Delta t})$.

\subsection{B. Effect of decoherence during the measurement process} 
Let us now consider the effect of decoherence during a single measurement only.  If we denote by $\rho_0$ the initial resonator density operator, the total state at the end of the Ramsey sequence is 
\begin{equation} 
\rho(\tau)= R_{\pi/2}(0) e^{\mathcal{L}\tau} \left(R_{\pi/2}(\varphi) \rho_0\otimes |g\rangle\langle g|  R^\dag_{\pi/2}(\varphi) \right)R_{\pi/2}^\dag(0).
\end{equation}Ê
For the expectation value at $t_1=t_0+\tau$ we obtain 
\begin{equation}
\langle Z(t_1)\rangle =  {\rm Re} \, {\rm Tr}Ê\{Êe^{i\varphi} \rho_{eg}(\tau)\}, 
\end{equation}
where $\rho_{eg}(t)= \langle e |e^{\mathcal{L}t}(\rho_0\otimes |e\rangle\langle g|) |g\rangle$ is the reduced operator describing the qubit coherence. 
According to the master equation~\eqref{eq:MasterEq} and Hamiltonian (1) in the main text this operator evolves as
\begin{equation}
\dot \rho_{eg}= -\left( i f_-(t)\Delta(t)+ \frac{1}{T_2} \right) \rho_{eg} -i \omega [a^\dag a ,\rho_{eg}] -i \lambda f_e(t) (a^\dag +a) \rho_{eg} + i\lambda f_g(t) \rho_{eg} (a^\dag +a)+ \mathcal{L}_m\rho_{eg}.
\end{equation} 
Equivalently, we can define the characteristic function $\chi_{eg}(\beta, t)=\text{Tr}\{\mathcal{D}(\beta)\rho_{eg}(t)\}$ and write 
\begin{equation}
\langle Z(t_1)\rangle  =  {\rm Re} \, \{Êe^{i\varphi} \chi_{eg}(\beta=0, t)\}.
\end{equation}
The evolution of the characteristic function is given by the Fokker-Planck equation 
\begin{equation}
\label{Eqcharacteg}
\begin{split}
\dot \chi_{eg}(\beta) = &
   i\omega \left(\beta \frac{\di}{\di \beta} -\beta^* \frac{\di}{\di \beta^*}\right) \chi_{eg}(\beta) - \frac{\Gamma}{2}\left(\beta \frac{\di}{\di \beta} + \beta^*\frac{\di}{\di \beta^*} \right)\chi_{eg}(\beta)  - \frac{\Gamma}{2}(2N+1) |\beta|^2 \chi_{eg}(\beta) \\
&-\left( i f_-(t)\Delta(t)+ \frac{1}{T_2} \right) \chi_{eg}(\beta)+ i\lambda f_+(t) \left(\frac{\beta+\beta^*}{2}\right)\chi_{eg}(\beta)  - i \lambda f_-(t) \left(\frac{\di}{\di \beta}-\frac{\di}{\di \beta^*}\right) \chi_{eg}(\beta).
\end{split}
\end{equation}
We solve this equation in three steps. First we make the ansatz 
\begin{equation}
\chi_{eg}(\beta, t)=  e^{-t/(T_2)} e^{i\phi(t)} e^{\beta \alpha_+^*(t)-\beta^*\alpha_+(t)} \chi_{I}(\beta,t),
\end{equation}
where 
\begin{eqnarray}
\dot \alpha_+&=& -(i\omega+\Gamma/2) \alpha_+ -i\lambda f_+(t)/2,
\end{eqnarray}
and
\begin{equation}
\dot \phi = -f_-(t)\Delta(t) - \lambda f_-(t)( \alpha_+(t)+ \alpha^*_+(t)).
\end{equation} 
For the remaining equation for $\chi_I(\beta,t)$ we introduce again $\Omega=\omega+i\Gamma/2$ and write it as 
\begin{equation}
\label{Eqcharacteg2}
\begin{split}
\dot \chi_{I}(\beta,t) = &\left[ i\left( \Omega \beta- \lambda f_-(t) \right)   \frac{\di}{\di \beta} -   i \left( \Omega^* \beta^*- \lambda f_-(t)\right)  \frac{\di}{\di \beta^*} - \frac{\Gamma}{2}(2N+1) |\beta|^2 \right] \chi_{I}(\beta,t).
\end{split}
\end{equation}
We now make the second ansatz
\begin{equation}
\chi_{I}(\beta, t)=  e^{-(N+\frac{1}{2})\left(|\beta|^2 - \beta \alpha_-^*(t)- \beta^*\alpha_-(t)  + \zeta(t)\right) } \chi_{II}(\beta,t),
\end{equation}
where 
\begin{equation}
\dot \alpha_- = -(i\omega+\Gamma/2)\alpha_-  - i \lambda f_-(t),
\end{equation} 
and
\begin{equation}
\dot \zeta =  - i \lambda  f_-(t)(\alpha_--\alpha_-^*).
\end{equation} 
This leaves us with the remaining equation for $\chi_{II}(\beta,t)$, which is given by
\begin{equation}
\label{Eqcharacteg3}
\begin{split}
\dot \chi_{II}(\beta,t) = &\left[ i\left( \Omega \beta- \lambda f_-(t) \right)   \frac{\di}{\di \beta} -   i \left( \Omega^* \beta^*- \lambda f_-(t)\right)  \frac{\di}{\di \beta^*} \right] \chi_{II}(\beta,t).
\end{split}
\end{equation}
This equation is solved by any function of the form 
\begin{equation}
\chi_{II}(\beta,t)\equiv \chi_{II}\left( x= e^{i\Omega t} \beta - i\lambda  \int_0^t e^{i\Omega s}Êf_-(s)ds\right),
\end{equation}
and the specific expression for $\chi_{II}(x)$ is determined by the initial conditions 
\begin{equation}
\chi_{II}(x)= e^{(2N+1)|x|^2/2}\chi_{eg}(x,t=0).
\end{equation}Ê
For an initial thermal state $\chi_{II}(x)=1$ and therefore 
\begin{equation}
\langle Z(t_1)\rangle  =  {\rm Re} \, \{Êe^{i\varphi} \chi_{eg}(\beta=0, t)\} = \cos\left( \varphi+\phi(t)\right) e^{-t/(T_2)} e^{-(N+\frac{1}{2})\zeta(t) }.
\end{equation}
For a static coupling and $\Gamma\ll\omega$ we obtain
\begin{equation}
\zeta(t)\simeq \frac{2\lambda^2}{\omega^2}\left[ (1-\cos(\omega t)e^{-\Gamma t/2}) + \frac{\Gamma t}{2}\right].
\end{equation} 
This shows that for $\lambda\sim \omega$ the signal of a single measurement decays with a total decoherence rate 
\begin{equation}
\Gamma_{\rm dec}= \frac{1}{T_2}+ \left(2N+1\right)\Gamma. 
\end{equation}
For $k_BT \gg \hbar \omega$ we obtain $N\Gamma\simeq k_B T/\hbar Q$, which then reproduces the decoherence time scales mentioned in the main text. Similar conclusions are obtain, when starting from a precooled state $\bar n\ll N$ or for $\lambda \ll \omega$, when a $\pi$-pulse sequence is obtained to amplify the displacement amplitude~\cite{BennettNJP2012}, but the results are rather lengthy and not discussed in detail here.

\end{document}